\documentclass[aps,prd,reprint,preprintnumbers,superscriptaddress,nofootinbib,longbibliography,floatfix]{revtex4-2}
\usepackage{amsmath}
\usepackage{graphicx}% Include figure files
\usepackage{dcolumn}% Align table columns on decimal point
\usepackage{bm}% bold math
\usepackage[colorlinks=true, citecolor=blue, linkcolor=blue, urlcolor=blue, linktocpage=true]{hyperref}
\makeatletter
\def\l@subsubsection#1#2{}
\makeatother

\newcommand{\omnifold}[1]{OmniFold}

\newcommand{\xtrue}{\ensuremath{\vec{x}_\text{true}}}

\begin{document}

% \preprint{APS/123-QED}

\title{A Practical Guide to Unbinned Unfolding}

\author{Florencia Canelli}\altaffiliation{CMS Collaboration}
\affiliation{Department of Physics, University of Zurich, 8006 Zürich, Switzerland}

\author{Kyle Cormier}\altaffiliation{CMS Collaboration}
\affiliation{Department of Physics, University of Zurich, 8006 Zürich, Switzerland}

\author{Andrew Cudd}\altaffiliation{T2K Collaboration}
\affiliation{Department of Physics, University of Colorado Boulder, Boulder, CO 80309, USA}

\author{Dag Gillberg}\altaffiliation{ATLAS Collaboration}
\affiliation{Department of Physics, Carleton University, Ottawa, ON K1S 5B6, Canada}

\author{Roger G. Huang}\altaffiliation{T2K Collaboration}
\affiliation{Physics Division, Lawrence Berkeley National Laboratory, Berkeley, CA 94720, USA}

\author{Weijie Jin}\altaffiliation{CMS Collaboration}
\affiliation{Department of Physics, University of Zurich, 8006 Zürich, Switzerland}

\author{Sookhyun Lee}\altaffiliation{LHCb Collaboration}
\affiliation{Department of Physics \& Astronomy, University of Tennessee at Knoxville, TN 37996, USA}

\author{Vinicius Mikuni}\altaffiliation{H1 Collaboration}
\affiliation{Physics Division, Lawrence Berkeley National Laboratory, Berkeley, CA 94720, USA}
\affiliation{National Energy Research Scientific Computing Center, Berkeley Lab, Berkeley, CA 94720, USA}

\author{Laura Miller}
\affiliation{TRIUMF, Vancouver, BC V6T 2A3, Canada}

\author{Benjamin Nachman}\altaffiliation{ATLAS \& H1 Collaborations}
\affiliation{Physics Division, Lawrence Berkeley National Laboratory, Berkeley, CA 94720, USA}
\affiliation{Department of Particle Physics and Astrophysics, Stanford University, Stanford, CA 94305, USA} \affiliation{Fundamental Physics Directorate, SLAC National Accelerator Laboratory, Menlo Park, CA 94025, USA}

\author{Jingjing Pan}\altaffiliation{ATLAS \& H1 Collaborations}
\affiliation{Physics Division, Lawrence Berkeley National Laboratory, Berkeley, CA 94720, USA}
\affiliation{Wright Laboratory, Yale University, New Haven, CT 06511, USA}

\author{Tanmay Pani}\altaffiliation{STAR Collaboration}
\affiliation{Department of Physics and Astronomy, Rutgers University, New Brunswick, NJ 08901, USA}

\author{Mariel Pettee}\altaffiliation{ATLAS Collaboration}
\affiliation{Physics Division, Lawrence Berkeley National Laboratory, Berkeley, CA 94720, USA}

\author{Youqi Song}\altaffiliation{STAR Collaboration}
\affiliation{Wright Laboratory, Yale University, New Haven, CT 06511, USA}

\author{Fernando Torales Acosta}\altaffiliation{H1 Collaboration}
\affiliation{Physics Division, Lawrence Berkeley National Laboratory, Berkeley, CA 94720, USA}

\date{\today}% It is always \today, today,

\begin{abstract}
Unfolding, in the context of high-energy particle physics, refers to the process of removing detector distortions in experimental data. The resulting unfolded measurements are straightforward to use for direct comparisons between experiments and a wide variety of theoretical predictions. For decades, popular unfolding strategies were designed to operate on data formatted as one or more binned histograms. In recent years, new strategies have emerged that use machine learning to unfold datasets in an unbinned manner, allowing for higher-dimensional analyses and more flexibility for current and future users of the unfolded data. This guide comprises recommendations and practical considerations from researchers across a number of major particle physics experiments who have recently put these techniques into practice on real data.
% \begin{description}
% Secondary publications and information retrieval purposes.
% You may use the \texttt{description} environment to structure your abstract;
% use the optional argument of the \verb+\item+ command to give the category of each item. 
% \end{description}
\end{abstract}

\maketitle

\tableofcontents

\section{Introduction}
\label{sec:intro}
% - why do we need unfolding? 
% - how does binned unfolding work, traditionally? 
% - what is unbinned unfolding? 
% - what is this white paper for? 

Science is driven by the pursuit of deeper truths about our Universe, but scientists must extract these truths using datasets collected by instruments that carry inherent measurement errors and resolution limits. To assess the validity of a theory by directly comparing it with experimental data, scientists must therefore account for the expected changes that will occur when a pristine theoretical prediction is altered by the imperfections of the measurement device. If the physics process to be measured and the model to be tested are both known \textit{a priori}, the standard approach used by researchers is ``forward modeling'': taking a theoretical prediction and then modifying it using a carefully-constructed, high-accuracy detector simulation that requires expert knowledge of the detector geometry and its calibrations in order to directly compare the prediction with real data. Each time one wants to test a new theory or physics process, however, this time- and compute-intensive process must be fully re-run, making it impractical for testing multiple theories, particularly if a new hypothesis is developed long after the measurements have already taken place. 

As an alternative to this paradigm, a real-world dataset can instead be corrected to try to remove any detector distortions from the data, thereby resulting in measurements that can be directly compared between experiments and theoretical predictions. Such a dataset could then be tested against any number of theoretical hypotheses without having to repeat the detector simulation each time. The nuanced process of removing detector effects from data is widespread across many areas of science and can be variously referred to as ``deconvolution'', ``denoising'', ``reconstruction'', or ``inverse modeling''. In High-Energy Physics (HEP), it has historically been called ``unfolding''. 

For decades, unfolding in HEP has been performed on binned datasets of no more than a handful of variables at a time. Algorithms such as Iterative Bayesian Unfolding~\cite{ibu} frame the experimental process as a series of transformations including a detector response matrix applied to the ``true'' dataset, i.e.\ an observable prior to the interaction with the detector --- or, equivalently, what a detector with perfect resolution and detection efficiency would measure. The detector response matrix encodes bin-by-bin measurement effects such as reconstruction errors, energy smearing, and background noise. Unfolding, in this framing, means effectively inverting this detector response matrix so that it can be applied instead to experimental data and thereby produce a truth-level dataset. Approximating this inverted matrix can produce numerical instabilities that can yield unreliable results, particularly as the number of variables --- and therefore the number of independent degrees of freedom --- increases. As a result, unfolded results from traditional binned strategies are restricted to fixed histogram bins for only a few simultaneous observables measured at once.

Over the past few years, HEP researchers have proposed new unfolding techniques that use machine learning pipelines to perform unfolding in an unbinned manner~\cite{Glazov:2017vni,Arratia:2021otl,bunse2018unification,Ruhe2019MiningFS,Andreassen:2019cjw,Andreassen:2021zzk,Arratia:2022wny,Chan:2023tbf,Datta:2018mwd,Bellagente:2019uyp,Bellagente:2020piv,Vandegar:2020yvw,Howard:2021pos,Backes:2022vmn,Shmakov:2023kjj,Alghamdi:2023emm,Diefenbacher:2023wec}, allowing for entire Monte Carlo simulated datasets to be corrected event-by-event such that the entire dataset aligns with the target. In addition to the inherent flexibility of removing a predetermined choice of bins from a measurement, these methods use neural networks that can readily process high-dimensional inputs, meaning that the unfolding can be performed for dozens or more variables at once without significant computational overhead. One such method that has gained recent prominence in the field is called OmniFold \cite{omnifold, Andreassen:2021zzk}. OmniFold is a density reweighting method that determines a set of weights that can transform a simulated dataset event-by-event to match the target, i.e.\ the data one would see with an idealized detector. These weights are derived from the ratio of the likelihoods of two samples. These ratios can be directly approximated by harnessing neural network classifiers, as these methods are known to naturally learn likelihood ratios in order to optimally discriminate between classes. 

Unbinned unfolding has now progressed well beyond simulation-only and proof-of-concept results. Between mid-2021 and mid-2025, a number of public results have made use of unbinned unfolding methods for measurements on datasets from at least five different HEP detectors: ATLAS, CMS, H1, LHCb, and STAR, as well as a study on highly realistic simulation from one accelerator-based neutrino experiment, T2K. This white paper is designed to further facilitate broader adoption of unbinned unfolding in the field by synthesizing the lessons learned from researchers involved in each of these measurements. In Section~\ref{sec:omnifold}, we summarize the core methodology of OmniFold, the unbinned unfolding technique that enabled each of these public measurements. In Section~\ref{sec:advice}, we summarize the specific choices made by the various analyses in areas including hyperparameter optimization, ensembling, uncertainty calculation, validation, and presentation of the final results. Finally, in Section~\ref{sec:conclusion}, we conclude and propose some future directions for unbinned unfolding in HEP using machine learning.

\section{Unbinned Unfolding}
\label{sec:omnifold}

\begin{figure*}[t!]
  \centering
  \includegraphics[width=\linewidth]{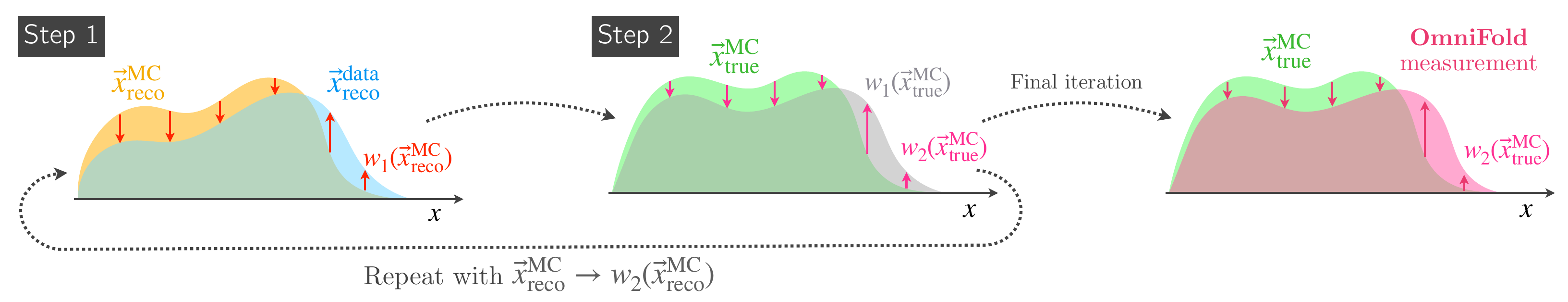}
  \caption{Illustration of the \omnifold{} method. In Step 1, MC is corrected to match data at the detector level, producing a weighting function $w_1(\vec{x}_{\text{reco}}^{\text{ MC}})$. In Step 2, a new function $w_2(\vec{x}_{\text{true}}^{\text{ MC}})$ is learned based only on particle-level quantities. The method proceeds iteratively, refining the function $w_2(\vec{x}_{\text{true}}^{\text{ MC}})$ such that the particle-level MC can be reweighted to give event yields and kinematics that match those observed in the data.
  \label{fig:omnifold}
}
\end{figure*}

Each unbinned analysis summarized in this work makes use of the classifier-based machine learning method called OmniFold~\cite{omnifold,Andreassen:2021zzk}, which is illustrated in Figure~\ref{fig:omnifold} and briefly reviewed here. OmniFold requires two data samples that provide distributions of three different sets of observables in total: 

\begin{enumerate}
    \item $\vec{x}^{\text{ MC}}_\text{true}$: a Monte Carlo simulation sample containing a list of events with a number of ``truth-level'' observables, i.e.\ particle-level properties as they would be captured by an idealized detector;
    \item $\vec{x}^{\text{ MC}}_\text{reco}$: a Monte Carlo simulation sample containing that same list of events, but instead with ``reco-level'' observables, i.e.\ the properties of those same events as they would be captured by a realistic detector; and
    \item $\vec{x}^{\text{ data}}_\text{reco}$: the experimental data that, by definition, only has ``reco-level'' observables.
\end{enumerate}
Every dataset also specifies a ``weight'' for each event. For the Monte Carlo datasets $\vec{x}^{\text{ MC}}_\text{true}$ and $\vec{x}^{\text{ MC}}_\text{reco}$, these weights are generally given by the Monte Carlo generator, but for the real data $\vec{x}^{\text{ data}}_\text{reco}$ these weights are all set to 1. 
An event indexed $i$ in the MC sample can hence typically be written as $(w_i,\vec{x}^{\text{ MC}}_{\text{true},i},\vec{x}^{\text{ MC}}_{\text{reco},i})$, and the 
 task of OmniFold is to use these three different sets of event properties to adjust the weight $w_i$ for each MC event such that the weighted sample of $\vec{x}^{\text{ MC}}_\text{reco}$ agrees with data. At that point, the weighted sample of $\vec{x}^{\text{ MC}}_\text{true}$ will provide an approximation 
 (or measurement) of the target, i.e.\ $\vec{x}^{\text{ data}}_\text{true}$. OmniFold and other likelihood-based methods do not change the observables themselves in the dataset --- only the weights. This is in contrast to unfolding based on generative models that change both the dataset observables \emph{and} the weights.

OmniFold makes frequent use of a useful feature of neural network binary classifiers: their learned decision functions can be used to construct a smooth reweighting function between the two datasets that participate in the binary classification. This property holds for a broad range of loss functions (see e.g.\ Ref.~\cite{Rizvi_2024}), but the binary cross-entropy loss function is a popular choice. With this choice of loss function, a classifier trained to distinguish whether a set of events $\vec{x}$ are drawn from either probability distribution $p_A(\vec{x})$ or $p_B(\vec{x})$ will learn a function $f(\vec{x})$ that minimizes:

\begin{equation*}
\label{eq:loss}
 \mathcal{L}_{\text{BCE}}[f] = - \int d\vec{x} \left(p_A(\vec{x})\log(f(\vec{x})) + p_B(\vec{x}) \log(1-f(\vec{x}))\right)
\end{equation*}

\noindent The function that minimizes this expression will satisfy, for any small variation $\delta f(\vec{x})$:
\begin{align*}
\delta \mathcal{L}_{\text{BCE}} &= \int d\vec{x} \left( -\frac{p_A(\vec{x})}{f(\vec{x})} + \frac{p_B(\vec{x})}{1 - f(\vec{x})} \right) \delta f(\vec{x}) = 0,
\end{align*}
therefore the integrand must be identically zero:
$$-\frac{p_A(\vec{x})}{f(\vec{x})} + \frac{p_B(\vec{x})}{1 - f(\vec{x})} = 0.$$
Solving this equation, we find that we can reconstruct the likelihood ratio:
\begin{align*}
\frac{f(\vec{x})}{1-f(\vec{x})} = \frac{p_A(\vec{x})}{p_B(\vec{x})},
\end{align*}
\noindent This function of the neural network classifier $f(x)$ can then be used to reweight one sample to look like the other by adjusting the event weight of each data point appropriately. In practice, this method will asymptotically approximate the likelihood ratio $p_A(\vec{x})/p_B(\vec{x})$~\cite{Hastie:2001,Sugiyama:2012}.

Like some binned unfolding methods that have been widely used in HEP for decades, OmniFold operates through multiple iterations. In fact, when applied to binned data, OmniFold is mathematically equivalent to Iterative Bayesian Unfolding~\cite{ibu}. Each iteration of OmniFold consists of two steps both of which involve estimating a reweighting function using the procedure described above: 

\begin{itemize}
    \item \textbf{Step 1:} First, a classifier is trained to distinguish between distributions of $\vec{x}^{\text{ MC}}_\text{reco}$ and $\vec{x}^{\text{ data}}_\text{reco}$. This classifier is used to construct a reweighting function $w_1(\vec{x}^{\text{ MC}}_{\text{reco}})$ that can be applied to make $\vec{x}^{\text{ MC}}_{\text{reco}}$ match $\vec{x}^{\text{ data}}_{\text{reco}}$ at the detector level. 
    \item \textbf{Step 2:} Next, a new classifier is trained to distinguish between distributions of $\vec{x}^{\text{ MC}}_\text{true}$ and the adjusted events from the previous step taken at the particle level: $\vec{x}_\text{true}^\text{ MC}$ weighted by $w_1(\vec{x}_\text{reco}^\text{ MC})$. A new reweighting function $w_2(\vec{x}^{\text{ MC}}_{\text{true}})$ is learned that can be applied to make $\vec{x}^{\text{ MC}}_{\text{true}}$ statistically match the target $\vec{x}^{\text{ data}}_{\text{true}}$ at the particle level. 
\end{itemize}

\noindent These two steps are then repeated for multiple iterations, where the next iteration instead takes $\vec{x}_\text{reco}^\text{ MC}$ weighted by $w_2(\vec{x}_\text{true}^\text{ MC})$
 as its input to compare with data. The updated reweighting function becomes a product of the previous one until a predefined number of iterations are performed and the method stops. 

\section{Practical Considerations}
\label{sec:advice}

\subsection{Experimental Overview}

Table~\ref{table:overview} summarizes the eleven analyses considered.

\begin{table*}[ht!]
    \centering
    \begin{tabular}{ |c|c|c|c|c|} 
     \hline
     Experiment &Paper Link&Dimensions&Final State&Momentum Selection  \\
     \hline\hline
     ATLAS \cite{atlas} & \href{https://arxiv.org/abs/2405.20041}{2405.20041} & 24 & $Z$+jets & $p_{\text{T}}^{\ell\ell} > 200$ GeV \\ \hline
    ATLAS \cite{atlas_track_functions} & \href{https://arxiv.org/abs/2502.02062}{2502.02062} & 6 & Dijets & $p_{\text{T}}^{j_1} > 240$ GeV \& $p_{\text{T}}^{j_1} < 1.5\  p_{\text{T}}^{j_2}$\\ \hline
     CMS \cite{cms} &  \href{https://arxiv.org/abs/2505.17850}{2505.17850} & 8 & Minimum bias & $> 2$ charged particles with $p_{\text{T}} > 0.5$ GeV \\ \hline
     H1 \cite{h1_ben}& \href{https://arxiv.org/abs/2108.12376}{2108.12376} & 8* & High $Q^2$ DIS & $Q^2 > 150$ GeV$^2$ \\ \hline
     H1 \cite{h1_vinny}& \href{https://arxiv.org/abs/2303.13620}{2303.13620} & 10 & High $Q^2$ DIS & $Q^2 > 150$ GeV$^2$\\ \hline
     H1 \cite{h1_fernando}& \href{https://arxiv.org/abs/2412.14092}{2412.14092} & 8* & High $Q^2$ DIS & $Q^2 > 150$ GeV$^2$ \\ \hline
     H1 \cite{h1_full_phase_space}& \href{https://www-h1.desy.de/h1/www/publications/htmlsplit/H1prelim-25-031.long.html}{H1prelim-25-031} & Variable & High $Q^2$ DIS & $Q^2 > 150$ GeV$^2$ \\ \hline
     LHCb \cite {lhcb} & \href{https://arxiv.org/abs/2208.11691}{2208.11691} & 4 & $Z$+hadrons in jets & $20 < p_{\text{T}}^{j} < 100$ GeV and $p_{\text{T}}^{h} > 0.25$ GeV\\ \hline
     STAR \cite{star_jets}& \href{https://arxiv.org/abs/2307.07718}{2307.07718} & 6 & Jets &$20 < p_{\text{T}}^{j} < 50$ GeV\\ \hline
     STAR \cite{star_heavy_ions} & \href{https://arxiv.org/abs/2403.13921}{2403.13921} & 7 & Jets in heavy ions & $20 < p_{\text{T}}^{j} < 45$ GeV\\
     \hline
     T2K \cite{T2K} & \href{https://arxiv.org/abs/2504.06857}{2504.06857} & 6 & Muon + Proton & $p^p > 450$ MeV for single transverse variables \\ \hline
    \end{tabular}
    \caption{An overview of the recently-published experimental results that use unbinned unfolding methods. For most results, the unfolded dimensionality is the same at reconstruction-level and at truth-level, but * indicates the analyses for which the unfolded dimensionality at reconstruction-level was the full phase space, but the truth-level unfolded result is 8-dimensional. Full details of the phase space for each measurement, including $\eta$ selections, are listed in the individual papers. \label{table:overview}}
    \end{table*}
    
\subsection{Hyperparameter Optimization}

The unbinned unfolding process includes a number of hyperparameters that affect either the core unfolding methodology (e.g.\ number of unfolding iterations) or the neural network training (e.g.\ batch size). The OmniFold codebase \cite{omnifoldcode} includes a suite of recommended default hyperparameters that were generally sufficient for some analyses. Other analyses chose to optimize hyperparameters based on the particular needs of their experiments. Those who did choose to perform a hyperparameter optimization generally tracked the performance of the model in unfolding with two different datasets of MC simulations.

\subsubsection{Unfolding Hyperparameters}
\paragraph{Number of iterations:} OmniFold \cite{omnifold}, like its binned analog IBU \cite{ibu}, is an iterative method. However, the number of iterations needed to see good performance can vary. To estimate this number, most analyses used dedicated studies with a known target to measure the convergence of the unfolded dataset with the target over a large number of iterations, sometimes reaching up to 70 or 100. In practice, far fewer iterations were used for the actual analysis result, typically around 5, due to limitations in the detector resolution. One notable exception is the T2K study \cite{T2K}, which preferred as many as 20 to 40 iterations. It also explored using per-event weight changes after each iteration to deal with the still open question of how to choose the appropriate number of iterations in an unbinned manner. The default choice in the OmniFold software is set to 3 iterations. Analyzers can also report the unfolding results for different numbers of iterations to assess the impact of this choice, as in Refs.~\cite{Komiske_2022, komiske_2022_6519307}.

\paragraph{Network initialization:} Most analyses chose to train the networks for each unfolding step from scratch with the onset of each new iteration, which is also the default choice in the OmniFold software. However, using a pre-trained network as a starting point for training subsequent steps can potentially reduce the overall training time and improve the stability of the result, particularly in cases for which there is a large imbalance in statistics between simulation and data \cite{h1_full_phase_space}, or where the available data statistics are just generally low \cite{T2K}. 

\paragraph{Step 2 task:} Step 2 of the unfolding procedure is designed to learn a function that maps the truth-level MC to the target of unfolded data. At each iteration, however, this task can be framed either as performing this mapping in full or in part. ``In part'', in this case, means learning to map from the previous iteration's reweighted MC to the target and then composing each of these mappings together after training to learn the full mapping from initial MC simulation to the target. The default behavior in OmniFold is to learn this mapping in full. In the CMS analysis \cite{cms}, researchers found that learning the mapping from the previous iteration's reweighted MC sample yielded worse performance and larger variance in the OmniFold weights, though this behavior has not yet been tested extensively.

\subsubsection{Neural Network Hyperparameters}
\paragraph{Network architecture:} The classifiers across the various analyses took the form of dense neural networks with a small number of of hidden layers---typically three, although some analyses also explored architectures with two or four layers. The T2K study~\cite{T2K} found measurably worse performance with smaller networks. Hidden layers mostly used the \texttt{ReLU} activation function and the final output layer used a sigmoid activation. Hidden layers generally had a size of $\mathcal{O}(100)$, e.g.\ 100 or 200 nodes. 
\paragraph{Batch size:} Larger batch sizes of $\mathcal{O}(10^3)$ events are preferable in order to maximize GPU usage and improve stability of the results. The default recommendation is 128, but in practice, different analyses selected batch sizes ranging from 1,028 \cite{cms} to 50,000 \cite{star_jets}. 

\paragraph{Number of training epochs:} Instead of using a fixed number of epochs, most analyses implemented early stopping using either a pre-determined patience of e.g.\ 10 epochs or a minimum $\Delta$, i.e.\ a lower bound on the change in the loss that counts as an improvement to the model. Early stopping was the most common form of regularization used, but a few analyses also employed dropout and batch normalization.  

\subsection{Preprocessing}

Data preprocessing can have major effects on the ultimate performance of a neural network training. Each analysis had to determine not only how to represent the features used as inputs, but also how to standardize the inputs and weights. 

\paragraph{Input features:} Many analyses used jet features represented as four-vectors, e.g. $p_T$, $\eta$, and $\phi$. In some cases, these features were represented relative to another reference frame such as the recoiling electron~\cite{h1_full_phase_space}. A few analyses chose to apply Graph Neural Networks (GNNs) in order to process the particles as 3D point clouds~\cite{h1_vinny, h1_full_phase_space}. The ATLAS $Z+$jets analysis~\cite{atlas} chose to represent $\phi$ coordinates as a combination of $\text{sin}(\phi)$ and $\text{cos}(\phi)$ after a dedicated study that suggested this representation was more effective at reducing discontinuities in the final results. In addition to raw input features, summary statistics such as moments and cumulants can also be successfully unfolded, even higher-order ones with relatively small values (e.g. $\mathcal{O}(10^{-5}$)) \cite{atlas_track_functions}. Inputs were generally standardized  using their $z$-scores, i.e. $z = (x - \mu) / \sigma$, before using them in the training process.

The T2K study \cite{T2K} identified two outgoing particles for each event, the muon and proton, and used the angles and forward momentum of these particles as inputs. However, in events for which there was no outgoing proton --- whether because it was not reconstructed or because no proton was kicked out at truth level --- placeholder values were used for the proton kinematics. Additionally, this analysis separated events into different samples in detector space based on the locations within the detector that the particles were reconstructed, and separated them into different reaction topologies in truth space. This qualitative information was provided as one-hot encoded input to the neural network.

\paragraph{Negative weights:} Simulated samples of high-energy physics data can sometimes be associated with MC weights representing wide magnitude ranges, even including negative values, which can introduce difficulties when using these weighted samples in training neural networks. In the ATLAS $Z$+jets measurement~\cite{atlas}, several of the input MC samples suffered from a significant fraction of negative event weights and large spread of weights. For these samples, the original MC weight was replaced by a new weight following a procedure similar to the method described in Ref.~\cite{Nachman_2020}. The resulting samples had positive weights with a reduced spread (standard deviation), yielding datasets that are statistically compatible to those produced using the original weights. 

\paragraph{Calculating the differential cross-section:} Most analyses performed the unfolding task on normalized samples such that the neural networks learned the relative shape differences between the datasets. This choice allows for the use of standardization in the weights themselves as well as the input features. The measured differential cross-sections can then estimated by scaling the weights based on a calculation using the efficiency and ofiducial factor from the nominal MC sample as well as the fiducial cross-section and number of data events: 

\begin{equation*}
\frac{\mathcal{L}\cdot \sigma_{\text{fiducial}}\cdot \epsilon} {f_{\text{fiducial}}} = n_{\text{data}}.
\end{equation*}
These corrections are implicitly accounted for if the unfolding is done without normalizing the samples, as was done in the T2K study \cite{T2K}. Alternatively, if the goal is to measure the shape of an observable, the measurement can be normalized again after unfolding such that the weighted area sums to 1.

\paragraph{Simultaneously unfolding event- and particle-level features:} Unfolded features can potentially include both aggregate and individual characteristics of event constituents, e.g.\ a jet and its constituent particles. In the LHCb measurements \cite{lhcb}, to account for this, an initial weight of unity for each particle was scaled down by the particle multiplicity within the jet to allow the network learn the substructure of jets and consistently normalize the synthetic data by the number of jets.

\subsection{Background and Acceptance Effects}

These analyses broadly benefited from a lack of major background sources that would require a detailed background subtraction process. However, their methodologies can in some cases be extended for analyses with nontrivial backgrounds. We note that ``backgrounds'' can comprise irreducible backgrounds, i.e.\ physics processes with the same final states as the signal, as well as reducible backgrounds, i.e.\ events that have been misclassified as signal events due to detector artifacts or other errors during particle reconstruction and identification.

\paragraph{Effects from irreducible backgrounds:} Backgrounds were estimated to be small or negligible for the analyses covered in this report, and were generally either ignored or were treated by assigning an overall uncertainty. For example, the ATLAS $Z+$jets measurement \cite{atlas} had a primary background of top events, but these were small enough ($<0.25\%$ of total events) to simply subtract them from a set of pseudodata, run the full unfolding procedure, and then take the difference between this and the nominal result as an uncertainty. The CMS measurement \cite{cms} targeted minimum bias proton-proton collisions, and the fiducial space for the measurement is as inclusive as possible. In the STAR heavy ions jets analysis~\cite{star_heavy_ions}, backgrounds were modeled using an embedding simulation to remove residual effects. To mitigate any discrepancies between data and MC due to choice of embedding strategy, an extra weight was then multiplied to the reco-level jets to make their multiplicity and luminosity distributions look similar to real minimum-bias data. For future users dealing with nontrivial irreducible backgrounds, we recommend incorporating these backgrounds into the initial MC dataset as MC events with negative weights such that in total, the initial MC dataset corresponds to data with background subtracted.

\paragraph{Acceptance and reducible background effects:} These effects emerge due to events that pass the truth-level selections but not reconstruction-level selections, or vice-versa. Generally, the former is the dominant acceptance effect due to tracking or trigger inefficiencies. The default OmniFold \cite{omnifold} procedure handles these events by assigning them the average weight in their region of phase space, which has been found to yield better convergence than directly including them in the unfolding procedure \cite{T2K}. Unfolding a larger fiducial region than the one reported in the final measurement can also help mitigate acceptance effects---for instance, the ATLAS $Z+$jets measurement~\cite{atlas} measures the phase space $p_T^{\ell\ell} > 190$ GeV, but only reports results with $p_T^{\ell\ell} > 200$ GeV in order to mitigate acceptance effects due to migrations across this threshold. 
%The experimental resolution of $p_T^{\ell\ell}$ is approximately XXX GeV.

In the LHCb measurements of charged particles in jets \cite{lhcb}, the track reconstruction efficiencies along with particle identification corrections were applied to each reconstructed particle in data being unfolded based on a look-up table prepared prior to unfolding.

In the STAR measurement of jet substructure variables \cite{star_jets}, this correction is instead done after the unfolding, and involves applying an efficiency correction by dividing the normalized unfolded dataset over the efficiency function bin-by-bin. Backgrounds due to misidentified jets were estimated using simulations and were used to construct initial weights for the data by subtracting the fake rates \cite{star_jets}. 

The minor background from the minimum-bias CMS measurement \cite{cms} came from fake tracks induced by detector noise or misidentification of the algorithms, which resulted in events that passed the detector-level selection but failed the truth-level selection. Acceptance effects came from tracking inefficiency corresponding to events that passed the truth-level selection but failed the detector-level selection. Both effects were taken into account by adding two steps\footnote{\footnotesize After Step 1, a Step 1b is added in which the weighting function $w_1(\vec{x}_{\text{reco}}^{\text{ MC}})$ is pulled back to the truth-level $w_\text{1b}(\vec{x}_{\text{true}}^{\text{ MC}})$ to correct the detector acceptance. $w_{\text{1b}}$ is derived by training a classifier at the truth level and can weight the original MC events passing the selections on both truth and detector levels to the same events with weights from Step 1. $w_{\text{1b}}(\vec{x}_{\text{true}}^{\text{ MC}})$ is applied to MC events not passing the truth-level selection. After Step 2, a Step 2b is added, in which the weighting function $w_2(\vec{x}_{\text{true}}^{\text{ MC}})$ is pushed to the detector-level $w_{\text{2b}}(\vec{x}_{\text{reco}}^{\text{ MC}})$ to correct the detector background. $w_{\text{2b}}$ is from training a classifier at the detector level and weights the original MC events passing both selections to the same events weighted by Step 2. Then $w_{\text{2b}}(\vec{x}_{\text{reco}}^{\text{ MC}})$ is applied to MC events not passing the detector-level selection.} to each iteration of the unfolding, as described in Ref.~\cite{Andreassen:2021zzk}.

The T2K study \cite{T2K} dealt with both significant inefficiencies from acceptance effects and significant backgrounds, which are common features of measurements in neutrino detectors due to their large volumes and relatively low fidelity. The backgrounds and effects from non-uniform detector acceptance were dealt with by splitting the detector-level events into distinct samples based on the topology with which they were reconstructed in the detector, including background-enhanced samples to constrain background contributions to the overall signal. This qualitative sample information was included as additional input to the neural network used in the unfolding. Additionally, results on single transverse variables were only reported for events with proton momentum $>450$ MeV, but the unfolding was performed over the entire available phase space without this restriction.

\subsection{Ensembling}
Most analyses found that ensembling improved the stability of the unfolded result. ``Ensembling'' refers to training multiple independent instances of the same model and then using the mean or median weight. In principle, this ensembling could occur at the level of binned observables, but ensembling the weights themselves facilitates publishing a fully unbinned dataset that does not depend on any particular choice of binning. The standard error on this quantity across the trained models is then included as a source of uncertainty. This uncertainty can be thought of as measuring the impact of the stochastic nature of the neural network training --- i.e. choice of random seed --- on the final unfolded result. The magnitude of this uncertainty is generally small ($<2\%$). Most analyses used some form of ensembling, with numbers of ensembles typically ranging from 4 to 10. Both the ATLAS $Z+$jets measurement \cite{atlas} and the ATLAS jet track functions measurement \cite{atlas_track_functions} used an ensemble of 100 models, which helped to mitigate the compounding effects of these statistical fluctuations over a large number of uncertainties. Ensembling was usually performed by re-running the entire result, but as an alternative, the ensembling could instead be done independently for Step 1 and Step 2 of each unfolding iteration \cite{acosta2025stabilizingneurallikelihoodratio}.

\subsection{Uncertainties}
Despite the introduction of neural networks underlying the unfolding procedures in these analyses, the procedure of estimating uncertainties is, with a few exceptions, relatively standard for particle physics measurements. These uncertainty estimation strategies include bootstrapping for estimating statistical uncertainties and calculating the effects of varying the MC sample for estimating systematic uncertainties. One new uncertainty included in these measurements was an uncertainty associated with the neural network initialization, i.e. the inherent variation in repeating the neural network training with a different random seed but otherwise an identical configuration. These variations are typically small, but they affect every other reported uncertainty, so they are important to constrain. 

\subsubsection{ATLAS unfolding uncertainty}

In the ATLAS $Z+$jets measurement, the unfolding uncertainty consisted of two parts: an uncertainty due to the truth-level prior as well as an uncertainty due to the potential mismodeling of the detector effects \cite{atlas}. These effects were estimated separately and reported as two separate sources of unfolding uncertainty. Together, the unfolding uncertainty represented the dominant uncertainty across most of the measured phase space. 

The first component of the ATLAS unfolding uncertainty, which measures the sensitivity of the result to the modeling of the particle-level sample, is estimated via a data-driven process: the MC is reweighted at truth-level such that all of its reco-level observables agree with data, and the result of unfolding using this dataset is compared with the nominal unfolding result. Crucially, the reweighting function used here is a sequence of one-dimensional Gaussian-kernel functions, not the OmniFold procedure that is itself being probed. The second component of the ATLAS unfolding uncertainty, which measures the sensitivity of the result to potential mismodeling of the detector response for features not included in the unfolding, is sometimes called a ``hidden variable'' uncertainty. To estimate this component, a MC sample from a different generator is used in the unfolding, but it is first reweighted at truth-level to match the MC sample from the original generator as closely as possible. 

\subsubsection{CMS unfolding uncertainty}

In the CMS measurement~\cite{cms}, uncertainty templates are implemented as weighting functions applied to nominal MC events. These functions are learned by classifiers trained to distinguish nominal and alternative MC samples, either at truth- or reco-level, depending on the source of the uncertainty. The templates for the CMS unfolding bias, which originate from the mismodeled truth-level prior on the target variables to be unfolded, are constructed as weighting functions $w_\alpha(\vec{x}_\text{true}^\text{ MC})$ at the truth level from the nominal MC sample to the MC sample from each alternative model $\alpha$. When applied to the nominal MC events, $w_\alpha(\vec{x}_\text{true}^\text{ MC})$ makes their truth-level samples match the ones from the alternative model. Similarly, the templates for mismodeled detector responses, which come from the discrepancies between the reconstruction in the simulation and truth, are constructed as weighting functions $w_\text{det}(\vec{x}_\text{reco}^\text{ MC})$ at the detector level, mapping the nominal MC sample to the MC sample from the same truth-level model but alternative detector simulation.

A two-stage reweighting scheme handles mismodeled priors in kinematic variables not directly unfolded. First, a truth-level weighting $v^1_\alpha(\vec{x}_\text{true}^\text{ MC})$ is applied to MC events from the alternative model $\alpha$ to match their truth-level samples to the ones from the nominal model. Then, a joint weighting of both truth and detector level, $v_\alpha(\vec{x}_\text{true}^\text{ MC},\vec{x}_\text{reco}^\text{ MC})$, is applied to the nominal MC events to match their samples at both levels to the ones from the alternative samples weighted by $v^1_\alpha(\vec{x}_\text{true}^\text{ MC})$. The resulting weighting function keeps the truth-level of the nominal sample while mapping its detector response to that of the alternative model. These reweighted templates allow smooth interpolation between models via nuisance parameters and enable uncertainty estimation via bootstrapping.

\subsection{Validation}

Many analyses were performed ``blinded'', i.e.\ not considering the real data while configuring the analysis parameters. This meant using a different MC sample in place of the real data while developing the analysis, and then only using the real data after the analysis passed a number of predetermined validation criteria. Even if the analysis itself was not formally blinded, all analyses validated their unfolding procedure using alternate MC samples. 

\paragraph{Pseudodata:} The ATLAS $Z+$jets analysis \cite{atlas} constructed a realistic set of ``pseudodata'' in place of the actual data to validate their procedure before obtaining results using the data. The pseudodata has a known underlying dataset \xtrue{}, so the measured \omnifold{} result of any quantity can be compared to the desired ``target'' value. To produce the pseudodata, alternate signal and background MC samples were reweighted to match data using a dedicated reweighting algorithm. The full analysis was performed using this pseudodataset, and the unfolded results were evaluated by analyzing the $p$-value distributions between the unfolded result and the target truth pseudodata. The agreement was checked using not only the 24 input observables, but also on some select composite observables (i.e. observables that can be calculated using the 24 but are not included in the unfolding), for a series of 2D distributions and for a series of phase space cuts. 

The CMS analysis~\cite{cms} used a few sets of ``pseudodata" from detector simulations of alternative models that are not used in unfolding or uncertainty estimation. The pseudodata sets are generated from distinct generators, parton distribution functions, or underlying-event tunes. The unfolding results of pseudodata sets are compared with their corresponding truth-level samples. The agreements are used for hyperparameter optimization and have been confirmed for all the variables to be unfolded and their multi-dimensional distributions.

\paragraph{Validation:} Technical closure tests and stress tests served to validate the unfolding methodology itself using different sets of MC samples. In the ATLAS $Z$+jets measurement, the stress tests were performed using a single MC sample to unfold itself, but the half used for data had additional stress weights applied. Two tests were performed: one where the weights were based on a deterministic function of the observables, and another where the weights were based on a stochastic function of the observables. In the T2K study \cite{T2K}, the test was performed using a set of fake data with reweightings that were functions of non-observable parameters. This resulted in the unfolding needing to learn a reweighting that is not a simple function of the quantities made available to it -- a type of stress test that is frequently done for neutrino cross-section measurements.

In the CMS result, bias and coverage tests were performed to evaluate the bias of the unfolded results compared to the truth and to validate the coverage of the unfolding uncertainties. The tests were based on unfolding pseudodata using a frequentist framework. Multiple toy experiments of unfolding were performed, in which both the MC and pseudodata samples were randomly deviated from the original ones according to the considered uncertainties. The cumulative behaviors of these toy experiments indicate the fluctuations if the measurement is repeated several times. The bias was evaluated as the deviations between the central values from the toy experiments and the pseudodata truth, and the coverage was estimated as the frequencies that the toy experiment results are compatible with the pseudodata truth, considering the uncertainties. 

Furthermore, the CMS analysis performed a bottom-line test to compare the differences between the unfolded results and the truth-level nominal MC sample to the ones between the (pseudo)data and the detector-level nominal MC sample. Since the unfolded results cannot be more discriminative to the models than the detector-level data, the former should always be smaller than the latter ones. A failure of the bottom-line test could come from the algorithm attributing the data-MC discrepancies from statistical fluctuations to the truth-level mismodeling and overfitting the given samples. %In case of a failure, the OmniFold hyperparameters should be adjusted to apply a stronger regularization. For example, one could use smaller neural networks or lower number of iterations.

\subsection{Presentation of Results}
Some community-sourced guidelines have already proposed how to best present unbinned measurements~\cite{Arratia:2021otl}. Across all of the results presented in this paper, however, though the unfolding procedure itself is performed unbinned, the unfolded results are presented in a binned histogram format for all but the ATLAS $Z$+jets measurement, which was presented in an unbinned manner.

\paragraph{Binned format:} Measurements are presented as individual histograms corresponding to each unfolded dimension. The specifications and bin counts of these histograms can then be uploaded to HEPData.

\begin{figure*}[t!]
    \centering
    \includegraphics[width=\linewidth]{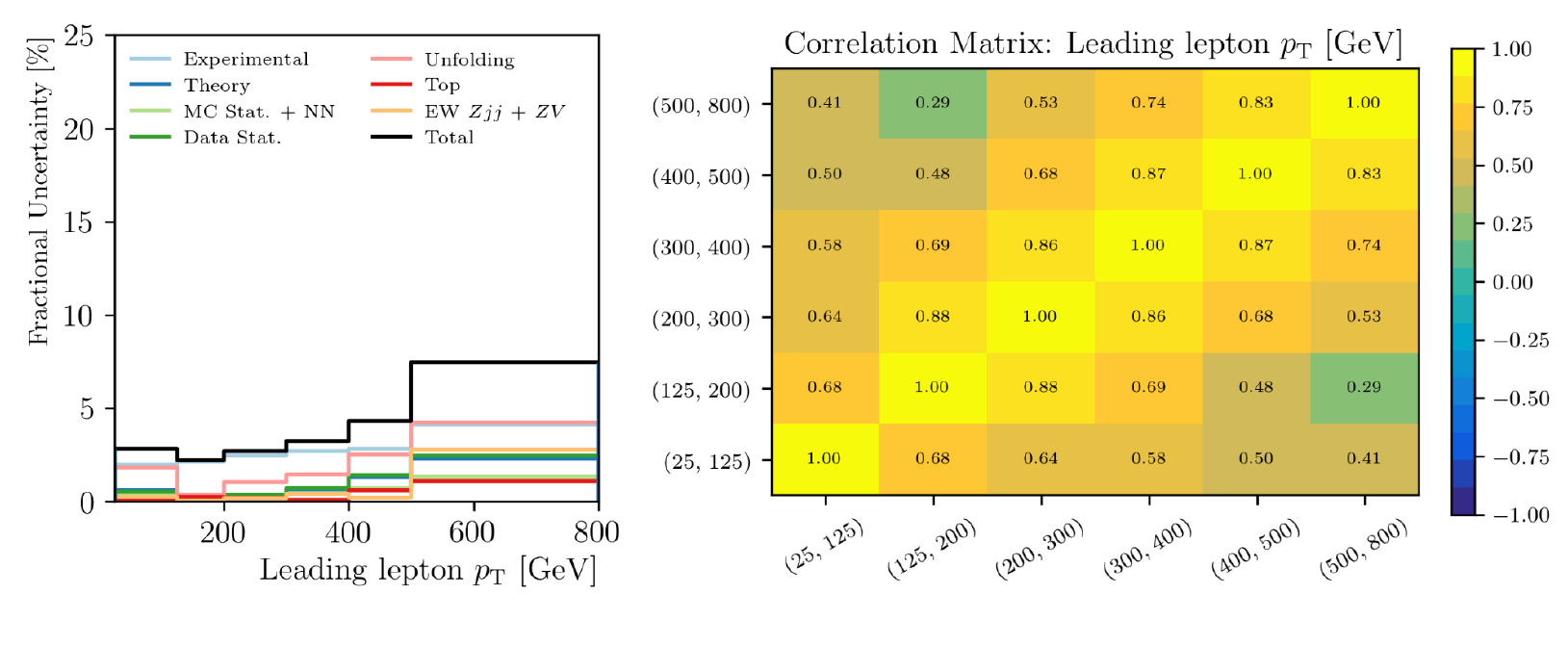}
    \caption{An example of the documentation from the ATLAS $Z+$jets measurement \cite{atlas} shows how to use the unbinned data to configure plots of not only the unfolded result, but also the uncertainty breakdown (left) and correlation matrix (right) using bins of one's choice for a given observable such as leading lepton $p_T$.}
    \label{fig:atlas}
\end{figure*}

\paragraph{Unbinned format:} In the ATLAS $Z+$jets measurement \cite{atlas}, the final result is presented as a set of Pandas DataFrames in which each DataFrame row corresponds with a truth-level Monte Carlo event, inspired by the example set by Ref.~\cite{komiske_2022_6519307} using CMS Open Data. More than one DataFrame is needed for the full ATLAS result in order to account for all sources of systematic uncertainty. Columns correspond with the 24 unfolded dimensions, i.e.\ truth-level particle quantities, along with hundreds of weights used to construct the nominal measurement and calculate uncertainties. These datasets are published on Zenodo \cite{atlas_zenodo} and are paired with a detailed codebase illustrating how to use them to replicate the results in the paper as well as how to produce new measurements (see Figure \ref{fig:atlas}). This presentation format was closely informed by the recent white paper that discusses the presentation of unbinned results \cite{Arratia:2021otl}. The CMS results will also be released in a similar unbinned format.

\subsection{Computing Requirements}
Across each of these measurements, an individual unfolding procedure could, in most cases, be run on a single NVIDIA A100 GPU with 40 GB or 80 GB of RAM in $\mathcal{O}(1$ hour), generally between 1 and 4 hours. The main factors influencing this training time include dimensionality of the unfolding, number of trainable model parameters, and number of unfolding iterations. If ensembling is used, identical copies of the unfolding setup are run up to $\mathcal{O}(100)$ times in order to reduce the uncertainty due to the stochastic nature of the training process. These ensembling runs are able to be fully trained in parallel, if compute availability permits. 

A full physics measurement, however, includes the calculation of a number of sources of uncertainty, and the number of uncertainties included can directly scale the computational resources required. Computational complexity for systematic uncertainties depends on the uncertainty sources and estimation strategies, requiring $\mathcal{O}$(100) unfolding runs if bootstrapping is used. Statistical uncertainties estimated through bootstrapping can also require about 50 to 100 individual bootstrap runs. 

In total, most analyses reporting significant computational needs for the unfolding portion of their experiments estimated that they used between approximately 500 and 10,000 GPU hours in total to perform their measurement once the full procedure had been fixed. Necessary computational resources depended heavily on the number of models used for ensembling and ranged from the STAR heavy ions analysis \cite{star_heavy_ions}, which did not use any ensembling and was able to run on a single GPU, to the 24-dimensional ATLAS $Z+$jets measurement \cite{atlas}, which used an ensemble of 100 models to calculate each weight and took approximately 25,000 GPU hours in total. 

\section{Conclusions and Outlook}
\label{sec:conclusion}
The recent emergence of more than 10 different high-energy physics publications using unbinned unfolding methods underscores that these methodologies are undoubtedly publication-ready. These results cover a variety of final states, six different experimental detectors, a range of unfolded dimensions (from four to 24 to the full phase space). Importantly, these methods have also been tested and validated in a number of different scenarios and have yielded unbinned cross-section measurements using large-scale physics datasets and full sets of uncertainties. As this field continues to expand, we expect that many more analyses will opt to both perform and publish their measurements in an unbinned fashion using the frameworks these first publications have established as a guide. 

\subsection{Future Directions}
While unbinned unfolding has now been proven effective for a range of particle physics applications, there remain several interesting directions for the development of future work in this area, including:
\begin{itemize}
    \item \textbf{Pre-trained models:} Would fine-tuning pre-trained models instead of training from scratch each time help improve the stability of the result as well as save computational resources, and should this become the new default methodology? (See Ref.~\cite{acosta2025stabilizingneurallikelihoodratio} for some additional discussion.)
    \item \textbf{Statistical uncertainties:} Are there viable alternatives to bootstrapping, to reduce the computational requirements needed in calculating statistical uncertainties? How can we best capture the statistical uncertainty not only in terms of the limitation of the total number of data events, but also in terms of the resulting instability in the neural network training?
    \item \textbf{Unfolding after many iterations:} In some cases, unfolding performance suddenly plummeted after many $(\mathcal{O}(100))$ iterations. Though this behavior did not affect the quality of the results ultimately presented here, which all relied on small numbers of iterations (5 or fewer), this behavior could merit future study.
    \item \textbf{Unfolding the whole phase space:} While the methodology for unfolding variable-length events is well-established in theory, we are only recently starting to see examples of this put into practice (e.g.\ Ref.~\cite{h1_full_phase_space}). To what extent do these recommendations apply to unfolding the whole phase space? Some researchers observed that when unfolding the whole phase space with datasets that were too similar, the unfolding performance could oscillate with each iteration --- what is responsible for this behavior, and how could it be mitigated?
    \item \textbf{Goodness-of-fit tests for unbinned data:} How can we best measure closure with a known target in an unbinned manner, therefore bypassing the typical $\chi^2$ and $p$-value binned statistical tests? Existing metrics include the Wasserstein distance \cite{dobrushin} and permutation tests~\cite{Williams:2010vh}, among others, but many open questions still remain in this area.
    \item \textbf{Generative methods:} Each of these measurements has employed a version of OmniFold \cite{omnifold, Andreassen:2021zzk}. Could other ML-based unbinned unfolding methods, including generative methods, yield competitive results?
    \item \textbf{Nontrivial background subtraction:} While there are proposed methods for putting nontrivial background estimation into practice, there is a strong need for researchers to put these methods to the test and explore which are most effective. 
\end{itemize}

\section*{Acknowledgments}
BN, MP and RH are supported by the U.S. Department of Energy (DOE), Office of Science under contract number DE-AC02-05CH11231. YS is supported by the U.S. Department of Energy (DOE), Office of Science under contract number DE-SC0004168. AC is supported by the U.S. Department of Energy, Office of Science, Office of High Energy Physics, under contract number DE-SC0010005. WJ, KC and FC are supported by the Swiss National Foundation under grant number 200020\_20497. TP is supported by the National Science Foundation under Grant number 1913624. We gratefully acknowledge the support of NSERC (Natural Sciences and Engineering Research Council of Canada). This work was also supported in part by the National Science Foundation under Grant No. 2311666.
% -----------------------------------------------------
\bibliography{bib, HEPML}

\end{document}